\documentstyle[prl,aps,twocolumn,epsf]{revtex}
\catcode`\@=11

\def\maketitle2{\par 
\begingroup
\let\cite\@bylinecite
\def\thefootnote{\fnsymbol{footnote}}%
\twocolumn[\@maketitle2\vskip2pc]%
\thispagestyle{plain}\@thanks
\endgroup
\def\thefootnote{\arabic{footnote}}%
\setcounter{footnote}{0}%
\let\maketitle2\relax \let\@maketitle2\relax
\let\@thanks\relax \let\@authoraddress\relax \let\@title\relax
\let\@date\relax \let\thanks\relax \let\@abstract\relax 
\let\@pacs\relax}

\def\abstract#1{\gdef\@abstract{{\par 
\bgroup
\ifdim\prevdepth=-1000pt \prevdepth0pt\fi
\hsize\columnwidth
\dimen0=-\prevdepth \advance\dimen0 by17.5pt \nointerlineskip
\small\vrule width 0pt height\dimen0 \relax}{~~}#1\egroup}}

\def\pacs#1{\gdef\@pacs{{\par 
\bgroup
\hsize\columnwidth \parindent0pt
\ifdim\prevdepth=-1000pt \prevdepth0pt\fi
\dimen0=-\prevdepth \advance\dimen0 by20pt\nointerlineskip
\egroup} PACS numbers:~#1}}

\def\@maketitle2{
\@preprint
\@title
\ifdim\prevdepth=-1000pt \prevdepth0pt\fi
\@authoraddress
\@date
\begin{list}{}{\leftmargin=0.10753\textwidth \rightmargin=\leftmargin
\itemsep=1pc\partopsep=-1pc}
\item\@abstract
\item\@pacs
\end{list}
}

\catcode`\@=12

\makeatletter

\def\compoundrel#1\over#2{\mathpalette\compoundreL{{#1}\over{#2}}}
\def\compoundreL#1#2{\compoundREL#1#2}
\def\compoundREL#1#2\over#3{\mathrel
  {\vcenter{\hbox{$\m@th\buildrel{#1#2}\over{#1#3}$}}}}
\makeatother

\begin{document}

\draft
\title{Quantum effects after decoherence in a quenched phase transition}
\author{Nuno D. Antunes$^1$, Fernando C. Lombardo$^2$, and Diana Monteoliva
$^3$}
\address{$^1$Centre for Theoretical Physics, University of Sussex,
Falmer, Brighton BN1 9QJ - UK}    
\address{$^2$Theoretical Physics Group, Blackett Laboratory, 
Imperial College, Prince Consort Road,
London SW7 2BZ - UK}
\address{$^3$Departamento de F\'\i sica, Facultad de Ciencias Exactas y 
Naturales\\Universidad de Buenos Aires - Ciudad Universitaria,
Pabell\' on I\\
1428 Buenos Aires, Argentina}


\abstract
{We study a quantum mechanical toy model that mimics some features of 
a quenched phase transition.
Both by virtue of a time-dependent Hamiltonian or by changing the
temperature of the bath we are able to show that even after
classicalization has been reached, the system may display
quantum behaviour again. We explain this behaviour in terms of simple 
non-linear analysis and estimate relevant time scales 
that match the results of numerical simulations of the master-equation.
This opens new possibilities both in the study of quantum effects in
non-equilibrium phase transitions and in general time-dependent 
problems where quantum effects may be relevant even after decoherence
has been completed.}

\date{\today}
\pacs{05.70.Fh,05.45.-a,03.65.Yz}

\maketitle2
\narrowtext

\section{Introduction}

The emergence of classical behaviour in quantum systems is a topic of 
great interest for both conceptual and experimental reasons \cite{exp}.
It is well established by now that the interaction between a quantum
system and an external environment can lead to its classicalization;
decoherence and the occurrence of classical correlations being the main 
features of this process (for a recent overview see \cite{Paz_Zurek}). 
     
A seemingly unrelated  physical problem where the interaction between a main
system and its surrounding environment is central is in determining
the dynamics of a phase transition. Usually, a
change in the properties of the system or the bath, 
forces the system to change phase via an out-equilibrium evolution.
It is natural to ask  what role decoherence plays in the
phase transition and conversely, how the time dependent nature of the
process affects the classicalization of the system.
     
In this article we explore two concurrent avenues. We look at what
may happen with the decoherence process when we have a time dependent 
setting (so far this problem has been mostly studied in kicked 
or driven systems; see for example Refs.\cite{BHJ} and \cite{MP}).
This is a very general question, and we use to guide us a 
simple toy model that naturally includes time dependent features.
This model also happens to mimic some properties of a non-equilibrium
second-order phase transition, giving us some clues as to what may
happen in a realistic case.

The paper is organised as follows. In the next section we introduce our
model and review the physical role of the different terms in the 
relevant evolution equations.
We describe how the `phase transition' is implemented and discuss 
estimates for the different time scales involved. 
In Section III we present the results of a series of numerical 
simulations for the evolution of an initial configuration of 
two de-localised Gaussian wave packets. This system is subject to a
sudden quench via an instantaneous change in the frequency sign.
Both the cases where the temperature of the environment is kept fixed
and allowed to change at the quench time are studied. We support the
numerical results with a detailed analytical analysis.
Section IV contains similar results this time taking as initial
condition a single Gaussian state centred at the global minimum.
In Section V we discuss the time dependent evolution 
of the linear entropy in the model, illustrating the loss of
purity of the system and clarifying the physical nature of the results
previously obtained.
Section VI contains final remarks and the main conclusions of
the paper.

\section{The model}

We will start by considering a quantum anharmonic oscillator
coupled to an environment composed of an infinite set of harmonic
oscillators. The total classical action for the system is given by:

\begin{eqnarray}S[x,q_n] &=& S[x] + S[q_n] + S_{\rm int}[x,q_n]\nonumber \\
&=& \int_0^t ds \left[{1\over{2}} M ({\dot x}^2 - \Omega_0^2(t) x^2 -
{\lambda\over{4}} x^4)\right. \nonumber \\
&+& \left. \sum_n {1\over{2}} m_n ({\dot q}_n^2 - \omega_n^2 q_n^2)\right] - 
\sum_n C_n x q_n,\end{eqnarray}  
where $x$ and $q_n$ are the coordinates of the particle and the oscillators 
respectively. The quantum anharmonic 
oscillator is coupled linearly to each oscillator in the bath with strength 
$C_n$.
This coupling leads to a simple quantum Brownian motion (QBM) 
model commonly used in the study of the quantum to classical transition 
\cite{QMB,hpz}. Tracing over the degrees of freedom of the environment one
obtains a master equation for the reduced density matrix of the system. 
From this one can derive the following evolution equation for the
corresponding  Wigner function \cite{Paz_Zurek}:

\begin{eqnarray}&&\dot{W_{\rm r}}(x,p,t)=\{H_{\rm syst},W_{\rm r}\}_{\rm PB} 
- {\lambda\over{4}} x 
\partial^3_{ppp}W_{\rm r}\nonumber \\
&+&2 \gamma (t) \partial_p(pW_{\rm r}) + D(t) \partial^2_{pp}W_{\rm r} 
- f(t) \partial^2_{px}W_{\rm r},
\label{fokker2}
\end{eqnarray}
where 

\begin{eqnarray}
\gamma(t) &=& -{1\over 2M\Omega_0}\int_0^t dt'\sinh(\Omega_0 t') 
\eta(t')\nonumber 
\\
D(t) &=& \int_0^t dt'\cosh(\Omega_0 t') \nu(t')\label{coef} \\
f(t) &=& -{1\over M\Omega_0}\int_0^t dt'\sinh(\Omega_0 t') \eta(t'),\nonumber
\end{eqnarray}
$\gamma (t)$ is the dissipation coefficient, $D(t)$ and 
$f(t)$ are 
the diffusion coefficients.
$\eta (t)$ and $\nu (t)$, the dissipation and noise kernels, 
are given respectively by:
\begin{eqnarray}\eta (t)& =& \int_0^\infty d\omega I(\omega ) \sin \omega t 
\nonumber \\
\nu (t) &=& \int_0^\infty d\omega I(\omega ) \coth {\beta
\omega\over{2}} \cos \omega t\nonumber, \nonumber\end{eqnarray} 
where $I(\omega )$ is the spectral density of the environment.

The first term on the right-hand side of Eq.(\ref{fokker2}) is the
Poisson bracket, corresponding to the usual classical evolution. The second 
term includes the quantum correction (we have set $\hbar = 1$). The last 
three terms describe dissipation and diffusion effects due to coupling to the 
environment. In order to simplify the problem, we consider a high-temperature 
ohmic ($I(\omega )\sim \omega$) environment. In this 
approximation the coefficients in 
Eq.(\ref{fokker2}) become constants: $\gamma (t) = \gamma_0$, $f\sim 1/T$,
and $D = 2 \gamma_0 k_{\rm B}T$. The normal diffusion coefficient $D$ is the 
term responsible for decoherence effects and at high temperatures is much
 larger than $\gamma_0$ and $f$. Therefore in Eq.(\ref{fokker2}), we may
 neglect the dissipation and the anomalous diffusion terms against the 
normal diffusion. It is important to note that 
the high-temperature approximation is well defined only after a time-scale 
of the 
order of $1/(k_{\rm B}T)\sim\gamma_0/D$ (with $\hbar = 1$). The relevant
period of evolution for our systems takes place at times comfortably 
larger than
this time-scale, safely in the validity regime of the approximation.

Time dependence will be introduced in the Hamiltonian by imposing a sudden 
change of sign of $\Omega_0^2$ (typical quench). 
This mass term is taken to be 
positive initially, the original symmetry being broken by $\Omega^2$ 
becoming negative.
 On a second stage we will also consider the case where the
temperature of the environment $T$ changes with time.
The change in the potential leads to the formation of degenerate
minima mimicking the breaking of symmetry in a second order-phase transition. 
In a realistic model one should address this problem in the context of 
quantum field theory \cite{lombmazzriv}. This is an extremely
difficult problem since
non-perturbative and non-Gaussian effects are relevant in the 
dynamical evolution of the order parameter undergoing the
transition and clearly numerical simulations are out of the question.
We trust that any non-trivial type of behaviour that may be a
feature of our simple quantum mechanical model will also be present
(and likely more strongly so) in the infinite dimensional case.

\section{De-localised initial states: quantum effects after 
decoherence}

We solve Eq.(\ref{fokker2}) numerically using a fourth-order spectral 
algorithm (numerical checks included carrying out simulations at different 
spatial and temporal resolutions). We chose $\lambda=0.1$, $D=0.3$ and set 
$\Omega_0^2=1.0$ initially. In order to understand the effects of the
change in the mass term on the decoherence process we look first at
the evolution of the quantum superposition of two Gaussian wave
packets:

\begin{equation}\Psi(x, t=0) = \Psi_1(x) + \Psi_2(x),\end{equation}
where
\begin{equation}\Psi_{1,2}(x) = N(t) \exp\left[-{(x\mp L_0)^2\over{2\delta^2}}
\right] ~ \exp{(\pm i P_0 x)}.\end{equation}
The initial 
$W_{\rm r}$ consists of two Gaussian peaks $W_{\rm r}^{1,2}$ separated by a 
distance $L_0$ (we chose $L_0=2.0$ and $P_0 = 0$) and an interference term 
$W_{\rm r}^{\rm int}$.
This quantum initial state
has been widely used in the literature to illustrate decoherence
phenomena (see \cite{Paz_Zurek} or \cite{hpz} for example) and its 
evolution will make clear the physical nature of the effects we will
observe. In the next Section we will chose a more realistic initial 
condition in terms of the dynamics of a phase transition.

 In order to 
visualise deviations from classicality effectively, we define the auxiliary 
quantity \cite{H}, 

\begin{equation}\Gamma(t) = \int dx\, dp \,\left[\vert W_{\rm r}\vert 
- W_{\rm
r}\right].\end{equation} 
 When the Wigner distribution is positive and possibly
identifiable with a classical probability distribution, $\Gamma$ is zero.
However if $\Gamma$ is positive  $W_{\rm r}$ must have negative values
due to quantum interference terms. We can thus use
positivity of the $\Gamma$ function as sufficient condition for 
non-classical behaviour.

\subsection{Mass quench at constant temperature}

\begin{figure} [hbt!]
\epsfxsize=8.6cm
\epsfbox{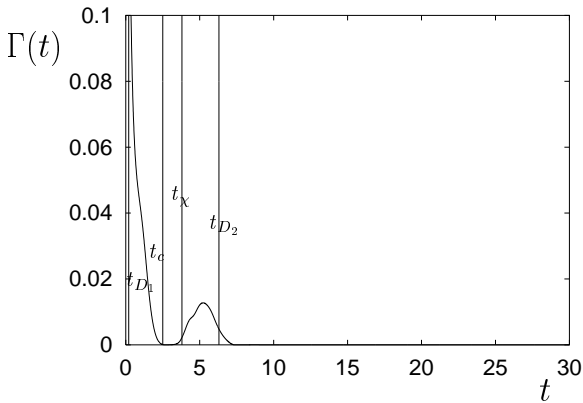} 
\caption{Evolution of $\Gamma$ when the potential changes its frequency 
from $\Omega_0^2 = 1 \rightarrow -1$. $D = 0.3$ and $\lambda = 0.1$.}
\label{figure1}
\end{figure}

We start the simulation by evolving $W_{\rm r}$ for some time with the 
positive mass
squared potential, in the presence of the bath.
During this period, the initial quantum interference terms are quickly
damped by the environment. Thus, for an early time $t_{\rm D_1}$, the system
decoheres and one is able to distinguish two classical probability 
distributions  
corresponding to the two initial Gaussian peaks evolving over phase 
space. Suddenly, at $t=t_{\rm c}$ we change
the frequency of the system from the initial positive
value $\Omega_0^2$ to a final  $-\Omega_0^2$.
The evolution picture changes dramatically when the frequency becomes
negative and instabilities are introduced in the system. 
In Fig.\ref{figure1} we can see the behaviour of $\Gamma$. Starting 
from a large initial value, $\Gamma$ quickly
tends to zero
as quantum fluctuations vanish and the systems becomes classical. The 
potential is quenched at $t_c$ and shortly after the system 
displays once again quantum behaviour for a period of time.

In order to understand this process we go back to early times, before
the quench. 
 From $t=0$ up to $t = t_{\rm c}$ the diffusion coefficient $D$ causes
the system to decohere, destroying quantum interference terms in a time
that can be estimated to be of the order of $t_{\rm D_1} \sim 1/(4L_0^2D)$,
where $L_0$ is the initial space separation between the peaks of the
Gaussian wave packets (see \cite{Paz_Zurek}). The normal 
diffusion term is dominant with respect to the quantum corrections, and 
thereafter the evolution is given essentially by the classical Fokker-Plank
flow.
For our choice of initial conditions we have $t_{\rm D_1} \sim 0.2$. This
is roughly the time quantum interference terms in the Wigner function
should fall to $1/e$ of their initial value (we have checked that this
is compatible with the decay of $\Gamma$ in the initial period of evolution
in our simulations). As soon as the frequency becomes negative, an unstable
point forms in the centre of the phase space with associated 
stable
and unstable directions. These are characterised by 
Lyapunov coefficients $\Lambda$ with negative and positive real parts
respectively \cite{ZP}. 

The new type of dynamics
gives rise to the possibility of squeezing along 
the stable direction.  
 The exponential stretching of the Gaussian 
packets in one of the directions due to the hyperbolic point is
compensated by an exponential squeezing. This will lead to a growth
of gradients in the Wigner function that will make the quantum
term in Eq.(\ref{fokker2}) comparable to the others. As a consequence
the system will be forced to explore the quantum regime again. In a more
quantitative fashion we have that the  
time dependence of the package width in the direction of the momenta
after the quench is given by  $\sigma_p(t) = \sigma_p(t_{\rm c})
\exp{[-\Lambda (t-t_{\rm c})]}$, where $\sigma_p(t_{\rm c})$ is the  
corresponding width at the time in which $\Omega_0$ changes sign.
From this we can estimate the $p-$derivatives of the 
Wigner function to grow as 
$\partial_p^n W_{\rm r} \propto \sigma_p^{-n}(t_{\rm c}) 
\exp{[n \Lambda (t-t_{\rm c})]} W_{\rm r}$. 
Clearly higher order derivatives grow
faster and at some point the quantum term with its third order derivative
will be of comparable magnitude to the classical terms in the Poisson
brackets (which are first order). This will happen (see \cite{ZP})
when the ratio  $\partial_p^3 W_{\rm r}/\partial_p W_{\rm r}$ becomes
of the order of $\chi^2 = \partial_xV(x)/\partial^3_{x}V(x) \sim 
\Omega_0^2/\lambda$ which characterises the scale of nonlinear 
terms. From this the time at which quantum effects become relevant is
calculated to be 
\begin{equation}t_{\chi} \sim t_{\rm c}+\Lambda^{-1}
 \ln [\chi\sigma_p(t_{\rm c}) ].\end{equation}
In the simulation used in our example we chose $t_{\rm c}=2.5$
(later than the time when the Wigner function becomes
definite positive).
 We evaluate $\chi \sim 3.2$ and numerically estimate 
$\sigma_p(t_{\rm c}) \sim 2.7$. We are also assuming the 
Lyapunov coefficient is given by the value corresponding to a linear 
potential $\Lambda = 2 \Omega_0^2 = 2.0$.
Therefore, the time in which quantum effects start being relevant  is 
given by $t_\chi \sim 3.8$. This is in good agreement with the time at which
the Wigner function displays negative values
once again, as can be seen in Fig.\ref{figure1}. 

From this point onwards quantum contributions increase, their growth
being limited by diffusion effects which limit
the squeezing of the Wigner function. The bound on the width of the 
packs is given by $\sigma_{\rm c}=\sqrt{2D/\Lambda}$
\cite{Paz_Zurek,ZP}. We use this to estimate
the second decoherence time scale. We assume that quantum effects become
maximal at a certain $t_{\rm max}$ (when in the numerical simulation
$\Gamma$ reaches its maximum) with a corresponding pack width
$\sigma_p(t_{\rm max})$ and that decoherence is effective after the
time when squeezing becomes of the order of the limiting value.
 This implies 
\begin{equation}t_{\rm D_2} = t_{\rm max}+ \Lambda^{-1} 
\ln [\sigma_p(t_{\rm max})/\sigma_{\rm c}],\end{equation}
which defines the decoherence time after the critical time. 
Using $\sigma_p(t_{\rm max})\sim 4$ and $\sigma_{\rm c}=0.5$ we obtain 
$t_{\rm D_2} \sim 6.3$, in reasonable agreement with the simulation
time for which quantum effects are exponentially suppressed 
(see Fig.\ref{figure1}).

\subsection{Mass quench with changing temperature}

\begin{figure}
\epsfxsize=8.6cm
\epsfbox{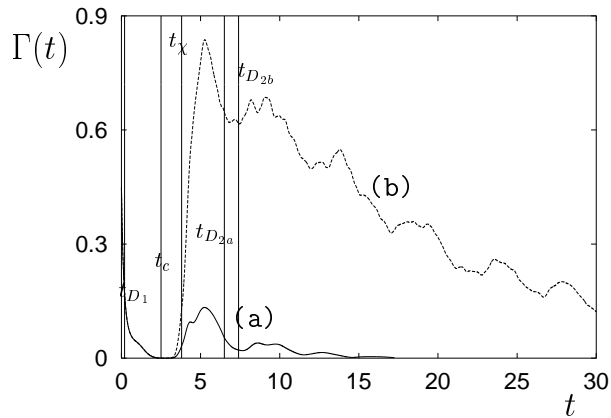}
\caption{Evolution of $\Gamma$ when the potential changes its frequency 
from $\Omega_0^2 = 1 \rightarrow -1$. A changing in the environment 
temperature is considered $D = 0.3 \rightarrow 0.1$ in curve (a) and $D = 0.3 
\rightarrow 0.003$ in (b).}
\label{figure2}
\end{figure}

The pattern of classical-quantum-classical behaviour found in the above
system with explicit time dependence is observed in
more generic situations. As a second example we have 
solved Eq.(\ref{fokker2}) allowing the bath temperature to decrease
simultaneously with the change in sign of the frequency term. These
conditions take us somehow closer to what would happen in a true 
second-order phase transition caused by a temperature quench. As a 
consequence, the diffusion coefficient, proportional to $T$, goes at
$t_{\rm c}$ from an initial high temperature value $D_0$ up to a final
lower value $D_{\rm f}$
(still in the high temperature regime
in order to ensure the validity of Eq.(\ref{fokker2})).
 In Fig.\ref{figure2} we see  the effect of changing the temperature
with  the classical potential (except for $D$ all simulations parameters
are the same as in Fig.\ref{figure1}). The analysis used in the previous 
example can be easily reproduced for this case. 
Both the initial decoherence time $t_{\rm D_1}$ and the time for the
re-introduction of the quantum fluctuations $t_{\chi}$ remain unchanged
as they do not depend on the temperature of the environment. 
The second decoherence time $t_{\rm D_2}$ is larger  
for a weaker diffusion term  (we have used $D_{\rm f} = 
0.1$ in Fig.\ref{figure2}-a and $D_{\rm f}=0.003$ in Fig.2-b). 
We have obtained  respectively $t_{\rm D_{2a}} \sim 6.5.$ 
and  $t_{\rm D_{2b}} \sim 7.4$. In the lowest 
temperature case 
(Fig.\ref{figure2}-b) the analytical prediction matches the
numerical result poorly. 
 This is due to the fact the estimation does not 
take into account the oscillations in the rate of decoherence 
coming from different orientations of the interference fringes when 
the Wigner function is moving around the unstable point. As the diffusion 
coefficient is smaller, the second decoherence time grows and the 
approximation of the upside-down potential in no longer valid.  
In any case, the analytic result can still be used as an estimated
lower limit for the second decoherence time. We have 
included it in our analysis in order to 
emphasise how dramatic the quantum effects are during the quenched 
transition.

\begin{figure}
\epsfxsize=8.6cm
\epsfbox{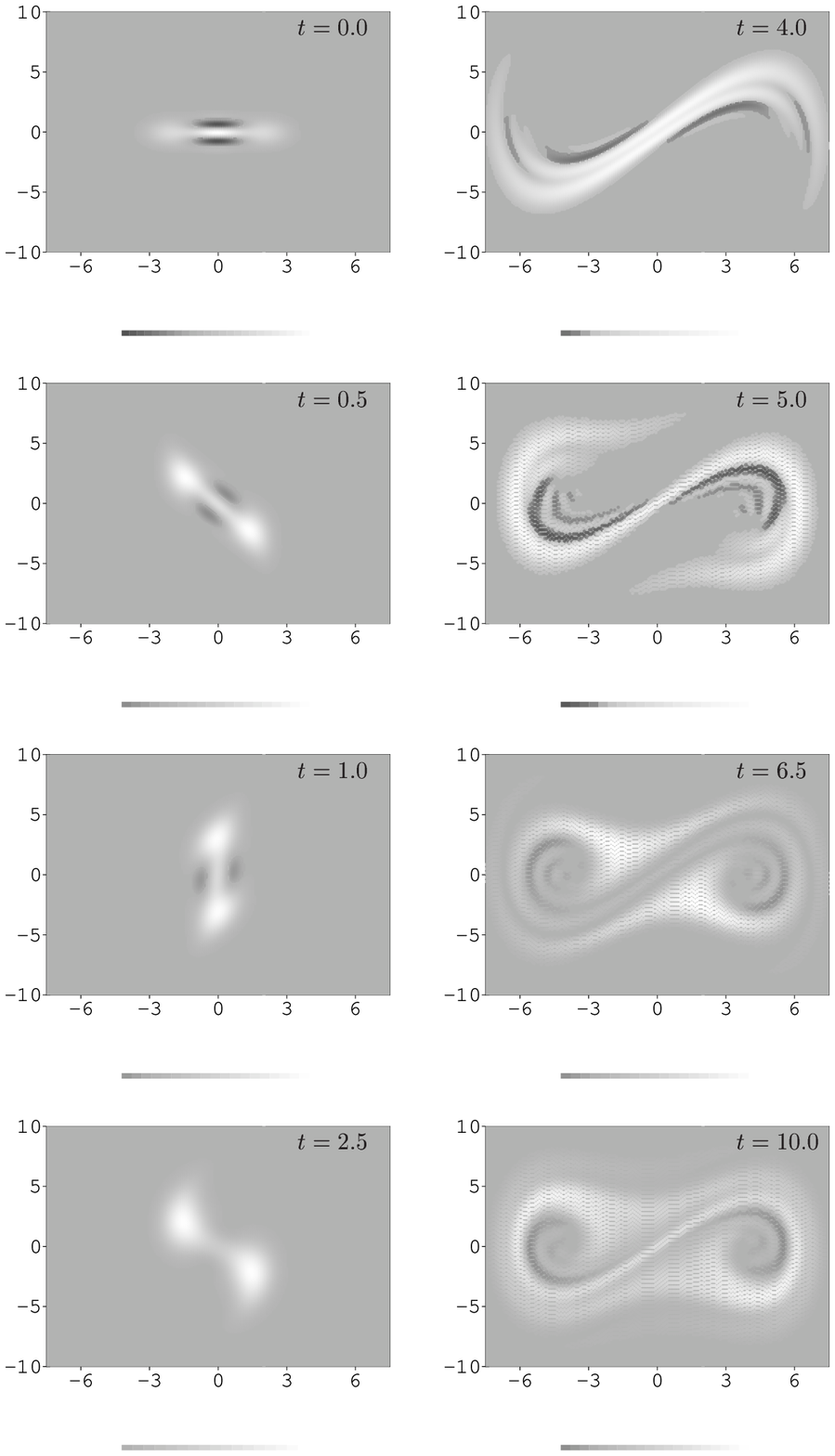}
\caption{Stroboscopic phase space for the evolution in
Fig.\ref{figure2}-a. Horizontal axis corresponds to $x$,
vertical axis to $p$. The medium grey shade on the background
corresponds to zero values for the Wigner function, lighter and
darker shades respectively to positive and negative values of $W(x,p)$}
\label{figure3}
\end{figure}

It is helpful to look at the Wigner function directly in order to
further clarify which regions of phase space are responsible for turning
$\Gamma$ positive. In 
Figs.\ref{figure3} we show $W(x,p,t)$ for the quench case corresponding  
to Fig.\ref{figure2}-a.

The four plots in the left column correspond to the decoherence period
before the quench. The two Gaussian peaks (light spots) rotate in
phase space around the minimum of the potential while the negative
components (dark patches) of the Wigner function are cleared away by
the environment. When the potential changes (right column) the wave
packets start spreading and exploring the new non-linear regions of
phase space giving rise to the dark interference patches. For longer
times decoherence takes over again and the Wigner function becomes once
more positively defined.
     
\section{Single initial Gaussian state}

\begin{figure}
\epsfxsize=8.6cm
\epsfbox{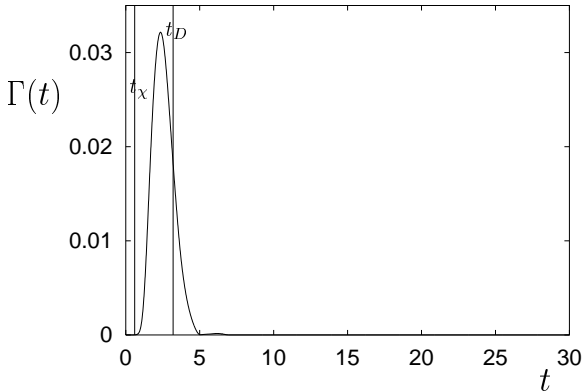}
\caption{Evolution of $\Gamma$ when the potential changes its frequency 
from $\Omega_0^2 = 1 \rightarrow -1$ for one Gaussian initially centred 
at $x=0$. ($\lambda = 0.1$).}
\label{figure4}
\end{figure}

\begin{figure}
\epsfxsize=8.6cm
\epsfbox{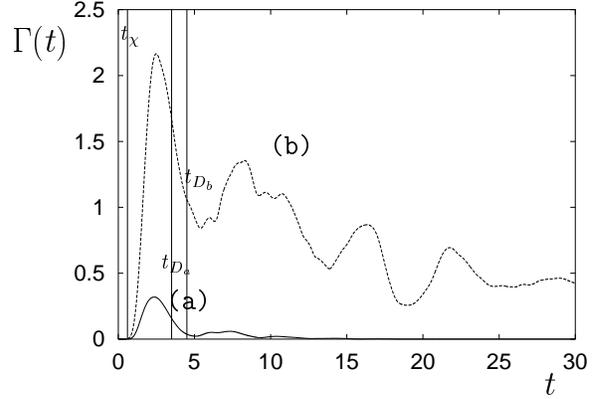}
\caption{Evolution of $\Gamma$ when the potential changes its frequency 
from $\Omega_0^2 = 1 \rightarrow -1$ for one Gaussian initially centred 
at $x=0$. ($\lambda = 0.1$). A changing in the environment temperature is 
considered $D=0.3 \rightarrow 0.1$ in (a) and $D=0.3 \rightarrow 0.003$ in 
(b).}
\label{figure5}
\end{figure}

As a further example we take a single Gaussian state centred at the 
global minimum of the quartic potential as initial condition. 
   This
is a more reasonable initial condition in terms of a realistic 
phase-transition, mimicking  a high-temperature thermal distribution.
It will also allow us to see that the above results are not an artifact
of the initial state. This initial 
Wigner function is already classical and so we ignore the initial
evolution period and take $t_{\rm c}=0$. Fig.\ref{figure4}
 and Fig.\ref{figure5} show
the $\Gamma$ function for the same quenches as before (without and with
temperature change respectively). The initial classical configuration
($\Gamma=0.$ for the initial time)  develops quantum effects 
as the classical potential and  
the temperature change. The relevant time scales are evaluated
as before and once again, the estimates are in good agreement with the
simulation results.
In the constant temperature case 
$\sigma_p(t_{\rm c}=0) \sim 0.7$ which gives $t_\chi \sim 0.6$ 
(see Fig.\ref{figure4}).
We also have  $\sigma_p(t_{\rm max}) \sim 3.2.$ and 
$\sigma_{\rm c}\sim 0.5$ leading to $t_{\rm D} \sim 3.2$,
which agrees with the numerical result.
     
Fig.\ref{figure5} shows the cases where the change in frequency is
followed by 
a change in the environmental temperature (same coefficients as in the 
example of Fig.\ref{figure2}). For Fig.\ref{figure5}-a 
 $\sigma_p(t_{\rm max}) \sim 3.3$ and $\sigma_{\rm c}= 0.3$, and 
therefore the decoherence time is $t_{\rm D} \sim 3.5$. This scale
is in good agreement with the numerical result. The estimation for
Fig.\ref{figure5}-b gives a 
decoherence time $t_{\rm D} \sim 4.5$ which again
 (as in the case of Fig.2-b) fails to fit the numerical result. We have 
included it in our analysis in order to 
emphasise how dramatic the quantum effects are during the quenched 
transition.

\section{Linear entropy}

One of the most salient features of the quantum to classical transition 
concerns the production of entropy as a consequence 
of the entangling interactions between the system and the environment.
 In order to clarify the nature of the post-decoherence quantum effects 
in the systems simulated above we have 
looked at the corresponding time evolution of the linear entropy.
This is given in terms of 
the density matrix  by \cite{MP}: 

\begin{equation}S_l(t)=-\ln [{\mbox{Tr}}[\rho^2(t)]].\end{equation} 

This quantity can
be easily obtained from the Wigner function giving a good measurement 
of the `loss of purity' of the system as it interacts with the
bath (see \cite{Paz_Zurek,MP}). 
We found that as expected the entropy increases throughout the whole 
evolution. The system starts as a pure state and while interacting
with the heat bath it looses coherence and simultaneously starts
behaving as a classical ensemble. When the potential changes it
evolves for some time as a 
quantum mixed system but the original `purity' is never recovered. In
this sense the decoherence process is irreversible.  
In terms of the Wigner function the linear entropy is related to 
the area of its non-zero component in phase space. Due to the coupling
to the environment the total area is not conserved, the Wigner
function keeps spreading at all times leading to permanent growth
of the entropy.
    
\begin{figure}
\epsfxsize=8.6cm
\epsfbox{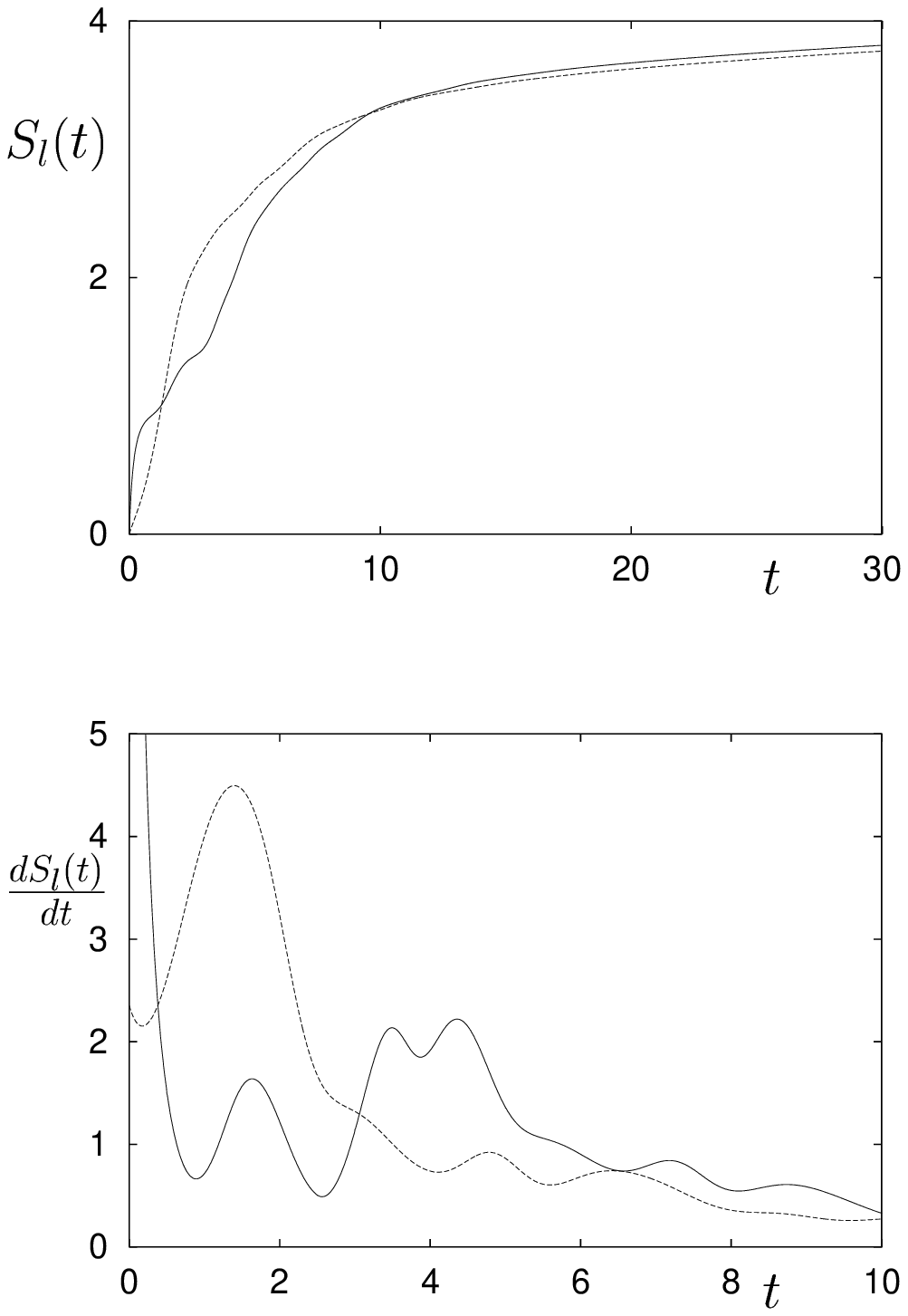}
\caption{In the top figure, we show the linear entropy for the de-localised 
(solid line) and single Gaussian (dashed line) initial states. In the 
bottom figure, we have the entropy production rate vs time  
for the same  initial conditions (solid and dashed lines as above).}
\label{figure6}
\end{figure}

Fig.\ref{figure6} shows the time dependent linear entropy (top plot)
and its production rate (bottom plot) for the two different initial
conditions considered before, in a quench with fixed environment 
temperature. 

In the case of the  double-gaussian initial state (solid line) there
is an initial period of evolution up to the first decoherence time 
$t_{D_1}=0.2$, where the linear 
entropy grows as a consequence of diffusion effects (as during the whole
evolution) and also due to the disapearence of initial interference 
terms which are washed away by the environment. As these vanish the
entropy production rate decreases as can be seen in the bottom plot.
 After an oscillation in the rate caused by the rotation 
of the Wigner function in the phase space which generates some low
amplitude interference terms, the rate reaches a minimal value near 
the quench time at $t=t_{\rm c}$. After $t_{\rm c}$, the entropy rate 
starts growing again as the system gets rid of the newly induced
interference terms. Finally at $t_{D_2}$, the entropy rate decreases 
to a low, slow decaying value driven by diffusion only.  

 The single gaussian evolution (dashed line) confirms this picture. 
From a low initial value (the initial state is free from negative 
terms) the entropy production rate grows as the quench generates 
interferences. Later, the environment cleans them out leading to the
final decaying rate. 

 We should stress that a growing linear entropy function does
not imply classicality (positivity of the Wigner function is an
extra necessary condition in order to have a classical probability 
distribution). Increase of $S_l$ tells us that 
the pure initial quantum state is evolving into a mixed state. 
It does not of course, tell us whether this mixed state is a classical
or quantum one.
In particular this is the case
between the quench time and the second decoherence time. During this
period quantum effects are re-introduced while the linear entropy is still
growing (faster even since its production rate increases).

\section{Final remarks}

We have shown, using an exact numerical evaluation of the Wigner function 
that quantum effects can be re-introduced after decoherence in several
systems with explicit time dependence. These quantum effects are
originated when the changing  dynamics introduce instabilities in
previously stable regions of the phase space. When this happens 
the dynamics of the Wigner function becomes more relevant than the
decoherence effects due to the environment (and the lowest the final
bath temperature the more dominant these are). The system then displays
quantum behaviour for a length of time until the environment manages to
catch up and force classicalization once again.

Since all examples so far were based on systems described by a double
well potential one could wonder whether our results could be a
consequence of possible tunnelling phenomena between the two minima.
Tunnelling is possible between symmetry related eigenstates with
energy below the barrier. The tunnelling time-scale for each pair is 
well known to be inversely proportional to the energy splitting of the
symmetry related pair of eigenstates. 
For the parameters of our system ($\Omega^2=1$, $\lambda=0.1$, $\hbar=1$)
only seven pairs of states are found below the barrier. Their energy 
splittings range from $\Delta E_0 \sim 10^{-12}$ to $\Delta E_7 
\sim 10^{-2}$, and thus the tunnelling would firstly be expected after 
$t \sim 100$. Therefore and considering the time scales in which our
simulations take place, tunnelling should play no role. The stretching and
folding of the Wigner function responsible for the observed effects
happens on both `sides' of the potential well independently. 
This is in agreement with the conclusion invariably found in the
literature (see for example Ref.\cite{tunel}) that tunnelling takes
place  rather slowly when compared with all natural time-scales in the
system.
     
\begin{figure}
\epsfxsize=8.6cm
\epsfbox{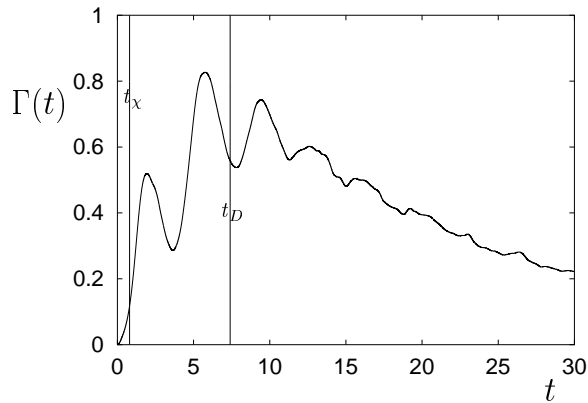}
\caption{Evolution for a single Gaussian packet cantered at $x_0=2$ ($t=0$) 
in a $x>0$ single-welled 
potential (and an infinite wall at $x=0$). We found $t_{\chi}= 0.8$ and 
$t_D= 7.4$ ($\Omega_0^2=-1$ and $D=0.01$). }
\label{figure7}
\end{figure}

 In order to confirm directly that tunnelling phenomena are not 
responsible for the effects observed, we solved 
numerically the problem of a single Gaussian packet cantered at $x_0 = 2$ 
evolving in the usual quartic potential but with its motion restricted to 
$x>0$. The 
resulting $\Gamma$ is shown in Fig.\ref{figure7}. As before quantum
behaviour is swiftly recovered, the corresponding time scales being in
good agreement with the analytical estimates.
     
Our results open up several interesting possibilities. The most
obvious one would be to try to `maximise' the recovering of quantum effects
to the extent of making them effectively permanent. An oscillatory frequency
\cite{ALM} that would continuously force instabilities into the system
could prevent classicalization or at least postpone it for a great length
of time. 
     
  In terms of the specific case of the dynamics of a second-order 
phase transition one could expect quantum effects to be present.
 Though the model used is a crude simplification of what happens in
a realistic phase transition, the same features of time-dependent
introduction of non-linearities would be present in that case,
leading to similar, probably stronger quantum effects. 
 Critical
properties of infinite dimensional systems  such as
critical slowing down could play an interesting role in the process.

\acknowledgments

We would like to thank S. Habib,  F. D. Mazzitelli, and J. P. Paz for 
comments and useful discussions. The work of N. D. A. was supported by a PPARC
Postdoctoral Fellowship and F. C. L. was supported by CONICET and
Fundaci\'on Antorchas.

\end{document}